\documentclass{WileyMSP-template}

\usepackage{multirow}

\begin{document}
\justifying

\title{Phase Transformation in Lithium Niobate-Lithium Tantalate Solid Solutions (LiNb$_{1-x}$Ta$_x$O$_3$)}

\maketitle


\author{Fatima {El Azzouzi*,}}
\author{Detlef Klimm,}
\author{Alexander Kapp,}
\author{Leonard M. Verhoff,}
\author{Nils A. Sch\"afer,}
\author{Steffen Ganschow,}
\author{Klaus-Dieter Becker,}
\author{Simone Sanna,}
\author{and Holger Fritze}



\begin{affiliations}

F. El Azzouzi and H. Fritze, {Institut f\"ur Energieforschung und Physikalische Technologien, Technische Universit\"at Clausthal, Am Stollen 19B, 38640 Goslar, Germany}, Email Address: fatima.ezzahrae.el.azzouzi@tu-clausthal.de

D. Klimm and S.Ganschow, {Leibniz-Institut f\"ur Kristallz\"uchtung, Max-Born-Str. 2, 12489 Berlin, Germany}

A. Kapp, L. M. Verhoff, N. Sch\"afer and S. Sanna, {Institut f\"ur Theoretische Physik, Justus-Liebig-Universit\"at Gie{\ss}en, Heinrich-Buff-Ring 16, 35392 Gie{\ss}en, Germany}

K-D. Becker, {Institut f\"ur Physikalische und Theoretische Chemie, Technische Universit\"at Braunschweig, 38023 Braunschweig, Germany}

\end{affiliations}



\begin{abstract}

The investigation of the structural phase transition in the vicinity of the Curie temperature 
$T_c$ of LiNb$_{1-x}$Ta$_x$O$_3$ crystals is motivated by the expected combination of 
advantageous high-temperature properties of LiNbO$_3$ and LiTaO$_3$, including high 
piezoelectric modules and remarkable high-temperature stability, respectively. $T_c$ marks 
the ultimate limit for exploiting the piezoelectric properties, however transition related 
structural modifications might impact this and other properties even below $T_c$.

Remarkably, the phase transition from the ferroelectric to the paraelectric phase, whose 
temperature strongly depends on the composition $x$, shows a significant drop in the 
activation energy of the electrical conductivity. The magnitude, temperature dependence 
and underlying mechanisms of this drop are discussed from a microscopic perspective. 
Molecular dynamics calculations in the framework of the density functional theory show that 
substantial displacements of the cations occur below $T_c$ for both the
end compounds LiNbO$_3$ and LiTaO$_3$, and might thus affect 
the electrical conductivity. Above $T_c$, the migration of lithium ions is presumably 
facilitated by a shortened diffusion path for the most favorable jump of the lithium 
ions. Electronic contributions to the conductivity, which become important above 900\,K,
are explained within the polaronic picture by the formation and migration of free 
small polarons.

\end{abstract}


\section{\label{sec:introduction}Introduction}

The single crystals of lithium niobate (LiNbO$_3$, LN) are well established in the field of optics. 
They are, for example, highly birefringent and exhibit high electro-optic coefficients. Moreover, 
the photoelastic as well as nonlinear optical properties are also distinctive \cite{Thierfelder10,Dues22}. 
These unique characteristics make LN an indispensable material for various applications, such as 
electro-optic modulation, frequency conversion, and optical parametric oscillation \cite{Volk08}. 
Similarly, lithium tantalate (LiTaO$_3$, LT) has also gained significant attention in various fields 
due to its large Non-Linear Optical (NLO) coefficients \cite {Miller1970,Dues22} and resistance to 
photorefractive damage \cite{Tangonan77}. Further, the piezoelectric properties of LN and LT render 
them ideal candidates for acoustic wave devices \cite{Zu2016,Koskela99}, holographic storage devices 
\cite{Imbrock99}, channel waveguides \cite{Kostritskii98,Kip98}, periodically poled structures 
\cite{Wong98}, and electro-optic deflectors \cite{Friel98}.

Crystals of LiNb$_{1-x}$Ta$_x$O$_3$ (LNT) solid solutions comprises the two border components LN and 
LT, which share the same crystal structure (space group $R3c$) but differ in their lattice and positional 
parameters, as well as their melting and Curie temperatures \cite{Roshchupkin2020,Xue2000}. The 
Czochralski method is commonly used for growing LN and LT crystals, which have a congruent composition 
containing approximately 48.4 and 48.5 mol\% Li$_2$O \cite{Bordui91,Malovichko99,Kostritskii09,Kim2001}, 
respectively. LN has a higher Curie temperature than LT. Also, LN exhibits higher piezoelectric 
coefficients but its high-temperature stability is poor due to its decomposition between 573\,K and 
1073\,K \cite{HornsteIner97}. On the other hand, LT crystals are shown to have better high-temperature 
stability which, for example, is connected with a lower electrical conductivity of LT above about 
873\,K \cite{Krampf2021}. However, the piezoelectric coefficients of LT crystals are lower than those 
of LN crystals \cite{Warren1970,Royer74,Campbell89}. To combine the superior properties of LN and LT, respectively, LNT crystals have been developed, with the aim of achieving 
both high piezoelectric coefficients and good high-temperature stability. This makes LNT crystals highly 
desirable for high-temperature applications and should allow for the fabrication of new devices including 
sensors for harsh environments.

The paraelectric to ferroelectric phase transitions of LN and LT are considered to be second-order 
phase transitions \cite{Sanna2017,Raeuber78}. In the case of LN, it occurs at fairly high temperature. 
Experimental data and theoretical modeling report transition temperatures for the congruent material 
in the range from 1413\,K to 1475\,K. A spontaneous polarization of 71\,$\mu$C/cm$^2$ is found at room 
temperature \cite{Volk08}. The structurally similar LT, however, has a lower phase transition temperature 
reported in the range from 874\,K to 958\,K \cite{Chen01,Nakamura08} and a spontaneous polarization of 
60\,$\mu$C/cm$^2$ \cite{Chen01}. The structural transition $T_c$ represents the upper limit at which 
devices exploiting the piezoelectric properties of the materials can be operated. However, as the phase
transformation is known to be a continuous process \cite{6306010,Phillpot04}, transition related
structural modifications might impact the materials properties even below $T_c$.

This study aims to uncover and explain changes in electrical conductivity that accompany the phase 
transition across various compositions. Our investigation involves experimental methods such as measuring 
electrical conductivity even above $T_c$, calorimetry, and theoretical exploration through \textit{ab initio} 
molecular dynamics calculations executed on large supercells.

\section{\label{sec:methods}Methodology}

\subsection{Crystal growth}

The LNT single crystals investigated here are grown at the Leibniz Institute for Crystal Growth in Berlin, Germany (IKZ), using the Czochralski method with induction heating. The starting materials are congruently melting LN and LT powders prepared from lithium carbonate (Li$_2$CO$_3$, Alfa Aesar, 5\,N), niobium pentoxide (Nb$_2$O$_5$, H.C. Stark, 4N5), and tantalum pentoxide (Ta$_2$O$_5$, Fox Chemicals, 4\,N), all dried before use. Two different LNT crystals are prepared by mixing LN and LT powders with LT mass fractions of 18\% and 34\%. Crystal compositions of $x \approx$ 0.42 and 0.64 result respectively. They are determined by averaging the distribution of Nb and Ta in the grown single crystals measured by Energy Dispersive Micro-X-ray Fluorescence ($\mu$-XRF) analysis in a pre-vacuum environment (20 mbar) using a Bruker M4 TORNADO spectrometer. Within the crystal sections used for the experiments discussed below, a variation in the composition of $\delta x = {0.02}$ is found. In the following, the term \textit{composition} referes exclusively to $x$ according to LiNb$_{1-x}$Ta$_x$O$_3$.

Seeds are pulled at a rate of 0.2 mm/h. The crystals grow along the [0001] direction (c-axis) in accordance with the orientation of the crystal seed. The process is carried out in a protective argon atmosphere with a small addition of oxygen (0.08 vol\%). Details can be found in \cite{Bashir23}.

To enable comparison of LNT with LN and LT,  congruent X-cut and Z-cut LN and LT wafers are purchased from Precision Micro-Optics Inc.\ (PMO, USA).

\subsection{Sample preparation}

The crystals are cut into rectangular plates of $5 \times 6$~mm$^2$. The only exception represent the samples with $x = 0.64$. They are slightly smaller and show an irregular shape.  To examine the electrical properties, platinum electrodes with a thickness of about 3~$\mu$m are deposited by screen printing on both surfaces of the samples using platinum paste (Ferro Corporation, No. 6412 0410). Subsequently, the samples are annealed at 1273~k for about an hour, with a heating rate of 2 K/min. Key information including sample name, composition, orientation, thickness and electrode size are given in Tab.~\ref{Tab:Samples}.  

\begin{table}
	\centering
	\caption{Overview of the samples used in this study. The composition $x$ refers to LiNb$_{1-x}$Ta$_x$O$_3$. The orientation denotes the crystallographic direction perpendicular to the largest surface.}	
	\begin{tabular}[htbp]{@{}llllll@{}}

		\hline
		\multicolumn{1}{l|}{Sample name}   & \multicolumn{1}{l|}{Composition $x$} & \multicolumn{1}{l|}{Manufacturer} & \multicolumn{1}{l|}{Orientation} & \multicolumn{1}{l|}{Thickness [mm]} & \multicolumn{1}{l|}{Electrode size [mm$^2$]}\\
		\hline
		LN-X & 0 & PMO & X & 0.5  & {5$\times$6}\\
		LN-Z & 0 & PMO & Z & 0.5 & 5$\times$6\\
		LNT42-Y & 0.42 & IKZ & Y & 0.68 & 5$\times$6\\
		LNT42-Z & 0.42 & IKZ & Z & 0.44 & 5$\times$6\\
		LNT64-Y & 0.64 & IKZ & Y & 0.73 & $\pi$$\times$3$^2$\\
		LNT64-Z & 0.64 & IKZ & Z & 0.76 & $\pi$$\times$2.5$^2$\\
		LT-X & 1 & PMO & X & 0.5 & 5$\times$6\\
		LT-Z & 1 & PMO & Z & 0.5 & 5$\times$6\\
		\hline
	\end{tabular}
	\label{Tab:Samples}
\end{table}

\subsection{Calorimetry}

Differential scanning calorimetry (DSC) is performed using a thermal analyzer (NETZSCH STA 449C F3, Germany). According to ASTM E1269, three consecutive measurements are carried out under identical conditions with empty reference and sample crucibles (baseline), Al$_{2}$O$_{3}$ powder is used as standard with known $C_p(T)$ function. Four subsequent heating runs are conducted up to 1463~K in a flow mixture of 40~ml/min Ar and 20~ml/min O$_{2}$, with isothermal segments for equilibration. The average $C_p(T)$ functions from the last three heating segments is used for further evaluation.

\subsection{Molecular dynamics}

Ab initio molecular dynamics as implemented in the Vienna Ab Initio Simulation 
Package (VASP) \cite{Kresse1993,Kresse1996,Kresse96_2} ist performed to model 
LiNbO$_3$ and LiTaO$_3$ within density functional theory at finite temperatures.
An canonical $NVT$ ensemble, controlled by the Nos\'e-Hover thermostat \cite{Nose84,Holian95}
was simulated at different temperatures to gather information about the temperature 
dependent lattice dynamics.
A Verlet algorithm with 2 fs time steps was used to integrate the equations of motion. 
A total simulation time of about 5 ps was considered. Due to the random initialization
of the velocities, an equilibration time of about 0.5-1\,ps is necessary, which is 
not considered in the evaluation of the AIMD trajectories. For the simulation of LN and
LT, large supercells consisting of 128 Li cations, 128 Nb or Ta cations and 384 O 
anions were used. Thereby, the thermal expansion as calculated within the quasi-harmonic
approximation is considered \cite{SimoMD23}. Due to the the supercell size in real space, 
the MD runs were performed on a single $k$-point, placed at the center of the Brillouin 
zone.
Projector augmented waves, calculated according to Bl\"ochl \cite{Bloechl94,Joubert1999} 
and implementing the PBEsol \cite{Perdew2008} formulation of the exchange-correlation 
potential are employed. The wave functions are expanded in a plane wave basis up to 
400\,eV.

From the AIMD runs, a real-time development of the polarization can be extrapolated. 
We employ thereby a simplified, approximated approach, in which the macroscopic 
polarization is defined as dipole moment per volume unit calculated with respect to 
a reference phase with $P_S=0$ (e.g., the paraelectric phase). More details about the 
calculation of $P_S$ are given elsewhere \cite{SimoMD23}.

\subsection{Electrical conductivity measurements}

To determine electrical conductivity, the samples are mounted in a furnace using an alumina support. Temperature control is realized using a type~B thermocouple located near the sample. Further, a Pt100 thermoresistor is used to compensate for the temperature fluctuations at the cold end of the thermocouple. The thermoelectric voltages are acquired by a digital voltmeter (Keithley-Instruments, USA). Measurements are conducted in air at atmospheric pressure up to 1623~K with a rate of 1~K/min. We perform short-term experiments in that sense that the dwell time at the maximum temperature is only one hour.

To determine the electrical impedance, the samples are connected by platinum foils and wires to an impedance/gain-phase analyzer (Solartron 1260, Ametek Scientific Instruments, Hampshire, UK). The impedance spectra are acquired in the frequency range from 1~Hz to 1~MHz with an excitation AC voltage of 50~mV. 

\begin{figure}[!h]
\centering
\includegraphics[width=0.6\linewidth]{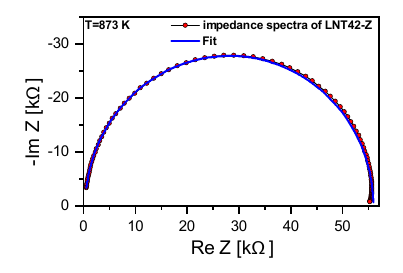}
\caption{Example of an impedance spectra of sample LNT42-Z at 873 k in the frequency range from 10~Hz to 1~MHz. Beside data, the fit of the equivalent circuit model (see text) is shown.}  
\label{fig:Impedanz}
\end{figure}

The electrical conductivity of the samples is described by an equivalent circuit (EC) model consisting of a bulk resistance (R$_{B}$) in parallel to a constant phase element (CPE), inferred from the obtained single semicircle in the complex plane. The bulk resistance is derived by least squares fitting of the related model parameters. Figure~\Ref{fig:Impedanz} shows exemplarily data at 873~k and the fitted model function. The intersection of the fitted semicircle with the real impedance axis is interpreted as bulk resistance $R_B$ and converted in the electrical conductivity $\sigma$(T) according to $\sigma = l/(AR_B)$, where $A$ and $l$ are the electrode area and the distance between the electrodes. Note that the uncertainty of the conductivity measurements amounts to up to 8\% \cite{Yakhnevych2023}. Up to about 1000\,K - 1230\,K (depending on the specific sample) the exponent of the CPE is found to be in the range 0.9 - 1.0. Above this temperatures, the exponent of the CPE is smaller than 0.9. Furthermore, the semicircles are less pronounced due to increasing inductive contributions. Nevertheless, the resistance can be determined as the intersection of fitted data and real impedance axis at the low-frequency side of the spectra.

The electrical conductivity $\sigma$ can be described by a Arrhenius relation as follows: 
\begin{equation}
	\label{eq:cond}
	\sigma=\sigma_0/T \exp \left(-E_A/(k_BT)\right)
\end{equation}
Here, $\sigma_0$, $E_A$, $k_B$ and $T$ represent a pre-exponential constant, the activation energy, the Boltzmann constant, and the absolute temperature, respectively. The factor $1/T$ reflects the correlation between mobility, relevant for electrical conductivity, and thermally activated diffusion according to the Nernst-Einstein relation.

\section{\label{sec:results}Results}
\subsection{Specific heat capacity and Curie temperature}
\begin{figure}
	\centering
	\includegraphics[width=0.6\linewidth]{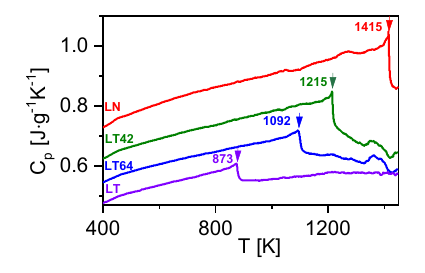}
	\caption{Experimental specific heat capacity of LiNb$_{1-x}$Ta$_x$O$_3$ measured in this study. The parameters indicate the Curie temperature of the respective sample.} \label{fig:Cp_LNT}	
\end{figure}

The DSC can be seen as a standard method to determine the specific heat capacity $C_p$. Here, the focus is on abrupt changes of $C_p$ at the phase transitions temperature only. Fig.~\ref{fig:Cp_LNT} depicts the $C_p$(T) curves for LN, LNT42, LNT64, and LT, which show the respective specific heat capacities at different temperatures for each material. The highest peaks in the curves indicate the $T_c$ of each material. Notably, LN exhibits the highest Tc value of 1415 K with LT42 and LT64 having Tc values of 1215\,K and 1092\,K, respectively. Lastly, LT has a $T_c$ value of 873 K.

\begin{figure}
		\centering
	\includegraphics[width=0.6\linewidth]{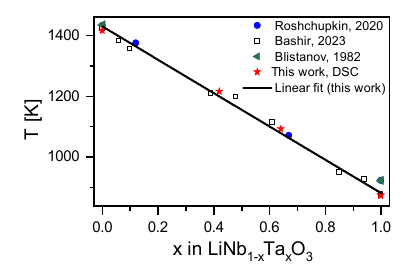}
	\caption{Curie temperature of LiNb$_{1-x}$Ta$_x$O$_3$ determined in this study, compared to
	literature data \cite{Roshchupkin2020} (Roshupkin, 2020), \cite{Bashir23} (Bashir, 2023) and
	\cite{Blistano82} (Blistanov, 1982).} \label{fig:Cp_Lit}
\end{figure}

Fig.~\ref{fig:Cp_Lit} shows the plot of $T_c$ as a function of composition, along with literature data. A linear correlation between Tc and Ta content is evident. Upon comparison with the literature data, it is observed that LN in references \cite{Roshchupkin2020}, \cite{Blistano82} and \cite{Bashir23} exhibits $T_c$ values of 1432 K, 1433 K, and 1423 K, respectively, while in this work, LN demonstrates a $T_c$ of 1430 K. Similarly, for LT, The literature value for $T_c$ are 923 K in references \cite{Roshchupkin2020} and \cite{Blistano82}, 879 K in reference \cite{Bashir23}, and 873 K in this work. Notably, LNT42 and LNT64 exhibit commendable linear agreement with the established literature data.

The temperature dependence of the $T_c$ data determined here can be approximated well by a linear function of the composition $x$ according to

\begin{equation}
\label{eq:Cp_lin}
T_c = \alpha + \beta x.
\end{equation}

with $\alpha =$ 1429~K and $\beta =$ -546~K. 

\subsection{Molecular dynamics}

\begin{figure}
\centering
\includegraphics[width=0.65\linewidth]{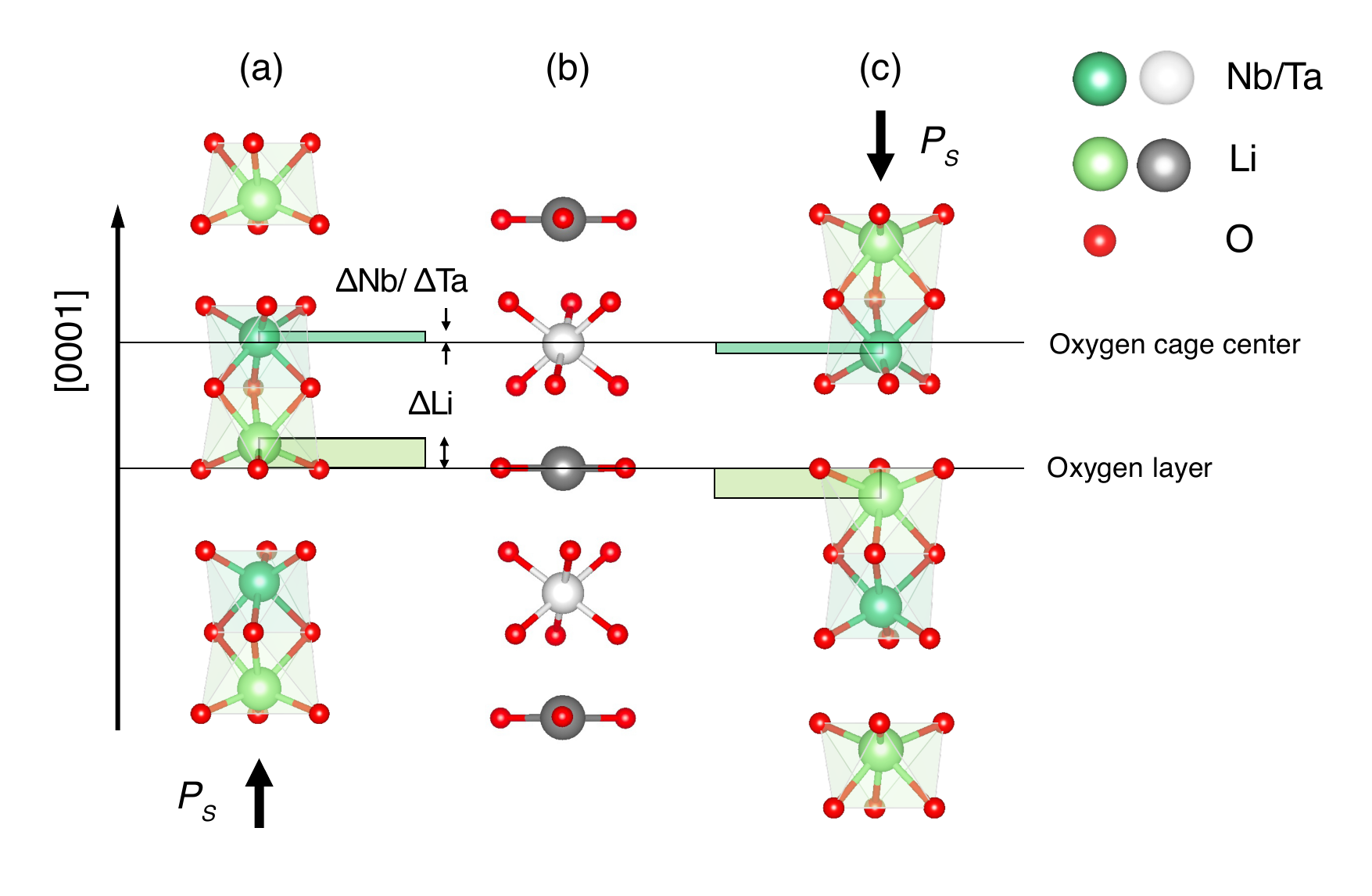}
\caption{Atomic structure of the ferroelectric (a, c) and paraelectric (b) phase of LiNbO$_3$ 
and LiTaO$_3$. The displacement of the Nb/Ta atoms from their positions in the paraelectric 
phase, i.e. from the center of the oxygen octahedra is denoted by $\Delta\mathrm{Nb}$ or 
$\Delta\mathrm{Ta}$, whereas the displacement of the Li ions from the oxygen planes is 
labeled by $\Delta\mathrm{Li}$.} \label{fig:Phasen}
\end{figure}

AIMD runs have been performed at temperature intervals of 100\,K from 300\,K to 
1100\,K for LiTaO$_3$ and from 900\,K to 1600\,K for LiNbO$_3$. 
From each MD trajectory, the displacements of Li and Nb/Ta from their equilibrium 
position in the paraelectric phase can be obtained. According to the $R\overline{3}C$ 
symmetry of the paraelectric phase, the equilibrium position of the Li ions is within 
a (0001) oxygen layer, while Ta/Nb ions are exactly at the center of the oxygen 
octahedra. This is shown schematically in Fig.\,\ref{fig:Phasen}.

\begin{figure}
\centering
\includegraphics[width=0.9\linewidth]{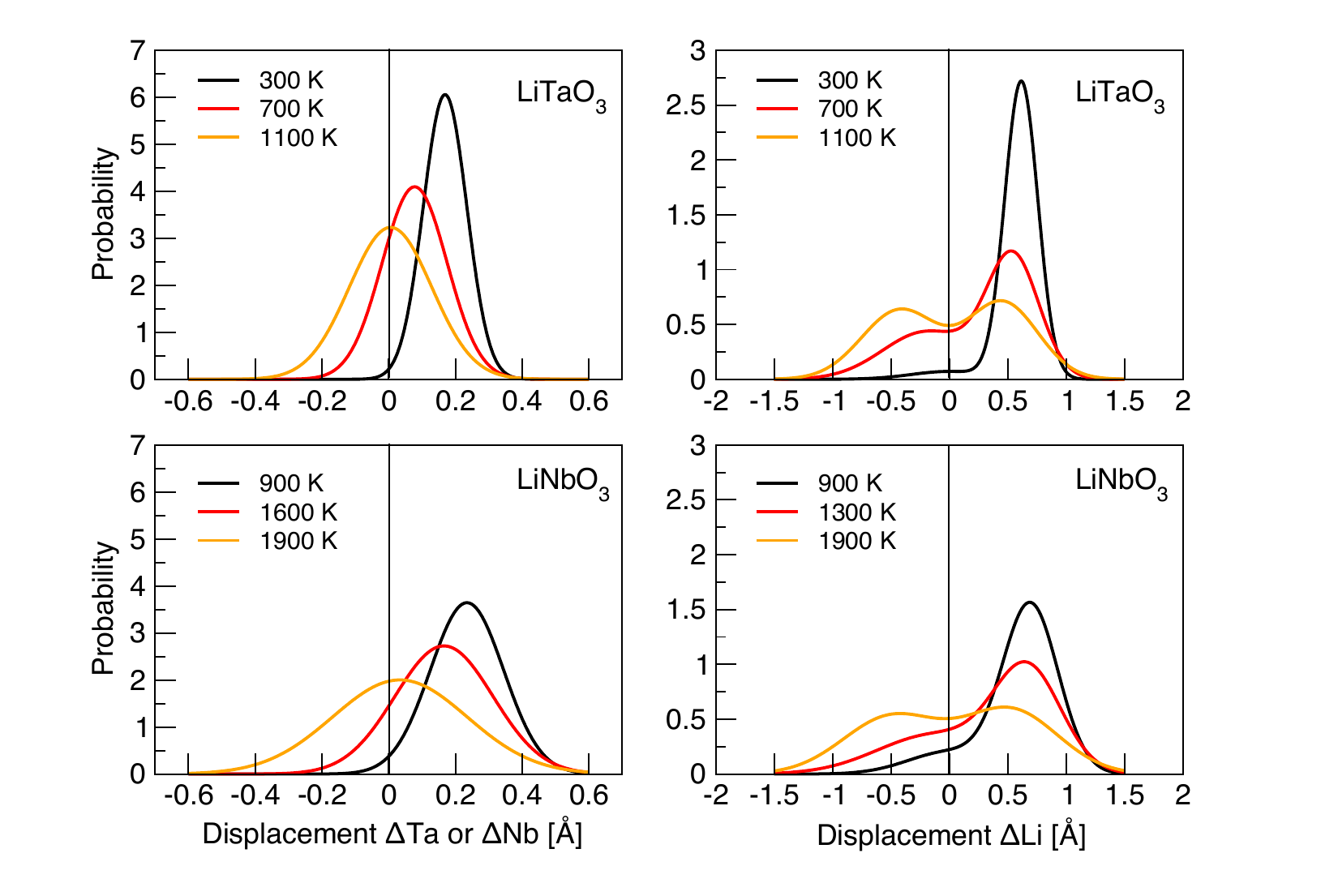}
\caption{Left column: Probability distribution of the parameter $\Delta$Nb (representing 
the displacement of the Nb ions from the center of the oxygen octahedra) at different temperatures
in LiTaO$_3$ and LiNbO$_3$. The distribution is unimodal at every temperature, as expected for 
a displacive phase transition. Right column: Temperature dependent probability distribution of the 
parameter $\Delta$Li (representing the displacement of the Li ions from the oxygen planes) in 
LiTaO$_3$ and LiNbO$_3$. The unimodal distribution well below the Curie temperature suggests
that all the Li ions are located above the oxygen planes. The symmetric, bimodal curve at above 
$T_c$ indicates that the Li ions are randomly distributed above or under the oxygen planes, as 
expected for an order-disorder phase transition.}
\label{fig:Li-Distr-Pos}
\end{figure}

The results from the AIMD are shown in Fig.\,\ref{fig:Li-Distr-Pos}. The displacement distribution of the Ta/Nb atoms as extrapolated from the
AIMD is well approximated by a single 
gaussian curve, centered at 0 within the paraelectric phase and continuously shifted to 
higher values when the temperature decreases. The displacements of the Ta/Nb 
ions are limited to ca. 0.35\,{\AA} from their equilibrium position, demonstrating that 
both Nb and Ta remain within the oxygen octahedra and thus suggesting a limited
mobility.

The displacement distribution of the Li ions reveals a striking different behavior, 
instead. At low temperatures, the Li ions are all above (or all below) an oxygen
(0001) layer. All the displacements are again well fitted by a single gaussian 
curve. When the temperature increases, a fraction of the Li atoms possesses enough
thermal enrgy to pass the oxygen layer. In the paraelectric phase, the Li atoms
oscillate around the oxygen layer, thus continously passing through the oxygen
(0001) plane. The corresponding displacement distribution is bimodal and can be
fitted by two gaussian curves.

On the one hand, this results in a dipole moment per unit cell that vanishes 
in average, explaining the absence of a spontaneous polarization at high temperatures.
On the other hand, this shows the high mobility of the Li ions, confirming the 
common assumption that they are the main mobile ionic charge carriers, and the main
carrier at all at moderate temperatures.

Interestingly, the AIMD shows that both for LiTaO$_3$ and LiNbO$_3$ substantial 
deviations from the equilibrium positions of the ferroelectric structure occur before 
$T_C$. In particular, Li ions display a non negligible mobility (a fraction of the
ions passing the oxygen plane) already a temperature well below $T_C$, which might
have an influence on the ionic conductivity.

\begin{figure}
	\includegraphics[width=0.6\linewidth]{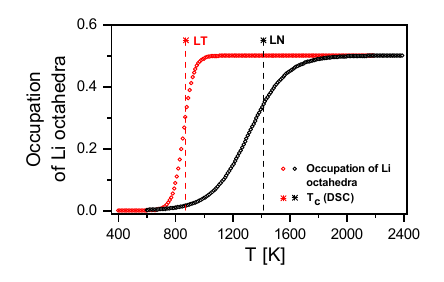}
	\centering
	\caption{Calculated temperature dependent occupation of the regular Li octahedra 
          (black triangles) and fit through a sigmoid function for LN and LT. Calculated 
          data is rigidly shifted to match the mesured $T_c$}
	\label{fig:Li-distr-Temp}
\end{figure}

In order to explore whether the Li mobility correlates with the electrical
conductivity, the occupation of the Li octahedra is shown in Fig.~\ref{fig:Li-distr-Temp} 
as a function of temperature. Details are discussed in Sec.\,\ref{sec:Discussion}.

For the interpretation of the AIMD results, a word of caution is in order. Atomistic 
calculations model ideal, stoichiometric materials. Real samples will feature a high 
density of point (intrinsic and extrinsic) and extended defects (dislocations, domain
boundaries, etc.). In particular, if the congruent composition is considered, the 
presence of a high concentration of Li vacancies must be taken into account, which 
can heavily affect the conductivity.

\subsection{Electrical conductivity}
\begin{figure}
	\includegraphics[width=0.6\linewidth]{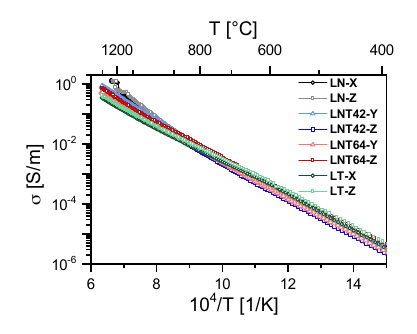}
	\centering
	\caption{Arrhenius diagram showing electrical conductivities of LN, LT, and LNT samples along different crystallographic directions.}
	\label{fig:conductivity}
\end{figure}

Figure~\ref{fig:conductivity} displays the electrical conductivity of LN and LT samples in 
both X and Z directions, as well as the Y and Z directions of LNT42 and LNT64 samples in air over a temperature range from 673\,K to 1503\,K in form of an Arrhenius representation. 
The data reveal a minor dependence of conductivity on composition up to about 873~K. Here, 
ionic conduction due to migration of lithium ions with an activation energy of about 1.3~eV 
dominates the electrical conductivity. The conclusions follows from the fact that lithium 
diffusion converted into electrical conductivity corresponds well to the electrical 
conductivity measured directly \cite{Chen07,Weidenfelder12,Kofahl23}.

\begin{figure}
	\includegraphics[width=0.6\linewidth]{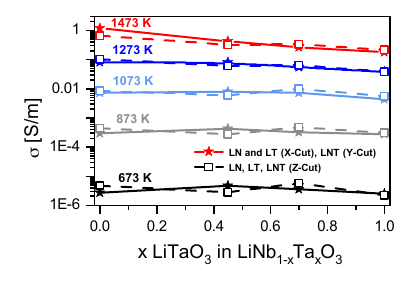}
	\centering
	\caption{Electrical conductivities of LN, LT, and LNT samples along the crystallographic X, Y, and Z directions as a function of LT concentration.} \label{fig:LT_Konz_Cond}
\end{figure}

To visualize this observation in more detail, Fig.~\Ref{fig:LT_Konz_Cond} shows the electrical 
conductivity as function of composition and crystallographic orientation for selected 
temperatures. Within the temperature range from 673~K to about 873~K, the composition has minor 
influence on electrical conductivity. Given that electrical conductivity measurements have an 
uncertainty up to 8\%, we conclude that the composition does not significantly affect the 
conductivity in this temperature range. However, above this temperature, an additional 
contribution potentially occurs with decreasing LT content, as suspected from Figs.~\ref{fig:conductivity} 
and \ref{fig:LT_Konz_Cond}. However, based on the previous presentations, the increasing 
conductivity and, in particular, the related temperature ranges are difficult to evaluate.

A more comprehensive picture is achieved by calculation of the logarithmic slope $E_{\sigma T}$ from Eq.~\ref{eq:cond}:

\begin{equation}
\label{eq:slope}
\ E_{\sigma T} = -k_B \frac{\partial \ln{(\sigma T)}}{\partial (1/T)}.
\end{equation}

If a single conduction mechanism dominates in a sufficiently large temperature range, 
$E_{\sigma T}$ is constant and can be assigned to an activation energy $E_A$. The approach, 
in turn, reveals distinct temperature ranges for the electrical conductivity that are not 
fully recognizable in the Arrhenius representation.

Figure~\ref{fig:Sigma_Tc}(a) shows the slopes $E_{\sigma T}$ of LNT for different 
compositions and crystal directions. A common observation is the independence of 
$E_{\sigma T}$ from the direction of the transport. Based on this, the simplified 
Fig.~\ref{fig:Sigma_Tc}(b) highlights characteristic features for the different 
compositions only. Additionally, Curie temperatures $T_c$ obtained from the specific 
heat capacity are indicated by vertical lines for the respective composition. A key 
observation is the abrupt decrease of $E_{\sigma T}$ when passing the transition 
from the ferroelectric to the paraelectric phase. In addition, four characteristic 
temperature ranges can be recognized, on which the aforementioned decrease of 
$E_{\sigma T}$ is superimposed:

\begin{figure}
	\centering
	\includegraphics[width=0.60\linewidth]{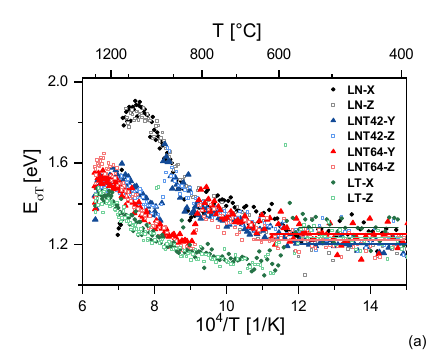}
	\includegraphics[width=0.6\linewidth]{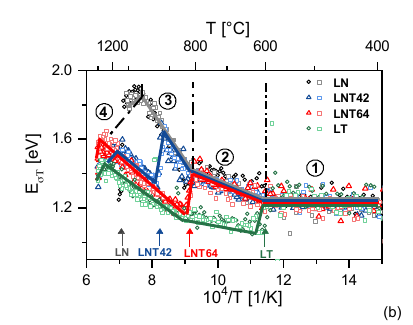}
	\caption{Logarithmic slopes of conductivity for representative LNT samples. Horizontal lines indicate approximately constant $E_{\sigma T}$ ranges (a). Scheme of the upper figure. The arrows indicate the Curie temperature for the respective composition while the dot-dash lines separate the temperature ranges discussed.}
	\label{fig:Sigma_Tc}
\end{figure}

\begin{itemize}
\item[(1)] At relatively low temperatures from 673~K to about 873~K, $E_{\sigma T}$ is 
constant and represents the activation energy $E_{AI}$ of the lithium migration mentioned 
above. Its values are determined by averaging $E_{\sigma T}$ and listed in Tab.~\ref{Tab:Ea} 
together with the associated specific temperature ranges. The values are represented as 
horizontal lines in Fig.~\ref{fig:Sigma_Tc}(a).

\item[(2)] With increasing temperature, a range with slightly increasing $E_{\sigma T}$ up 
to about  1073~K becomes visible. It is attributed to additional electronic conduction 
\cite{Yakhnevych2023}. Note that the slope of $E_{\sigma T}$ depends on the composition, 
as discussed in the context of Figs.~\ref{fig:conductivity} and \ref{fig:LT_Konz_Cond}. 
For $x$ = 1, a nearly constant slope follows, which enables to calculate the activation 
energies given in Tab.\ref{Tab:Ea}. Note, that the latter belong to the paraelectric 
phase of LT.  

\item[(3)] An even stronger increase in $E_{\sigma T}$ is observed above 1073~K. It 
extends up to about 1273~K and 1573~K for LN and LT respectively. The nature of this 
contribution is not yet known and subject of ongoing investigations. In the short-term 
experiments carried out here, degradation obviously plays a minor role, as the data for 
heating and cooling differ very little.

\item[(4)] Finally, a decrease of $E_{\sigma T}$ is observed with further increasing 
temperature. In case of LT, the decrease starts at about 1323~K and coincides with the 
drop at $T_c$. In case of the other compositions, the decrease starts at even higher 
temperatures, and is far off $T_c$.
\end{itemize}

\begin{table}
\centering
\caption{Activation energies of lithium ion conductivity calculated by averaging $E_{\sigma T}$ in the given temperature range.}	
\begin{tabular}[htbp]{@{}lll@{}}
\hline
\multicolumn{1}{l|}{Sample name} & \multicolumn{1}{l|}{Temperature range [K]} & Activation energy E$_{AI}$ [eV]\\
\hline
LT-X & 673 - 843 & 1.23\\
     & 878 - 1073 & 1.13\\
LT-Z & 673 - 843 & 1.28\\
     & 883 - 1073 & 1.12\\
\hline
LNT64-Y  & 673 - 853 & 1.24 \\
LNT64-Z & 673 - 883 & 1.22\\
\hline
LNT42-Y  & 673 - 853 & 1.20\\
LNT42-Z  & 420 - 923 & 1.25 \\
\hline
LN-X & 673 - 803 & 1.19\\
LN-Z & 673 - 803 & 1.20\\
\hline
\label{Tab:Ea}
	\end{tabular}
\end{table}

A common phenomenological observation in the vicinity of the phase transitions is 
an increase in $E_{\sigma T}$ with temperature, which is superimposed by an abrupt 
drop of $E_{\sigma T}$ at the phase transition temperature. This situation enables 
the extraction of the magnitude of the drop in $E_{\sigma T}$. In the vicinity, but 
not at the phase transistion temperature of a sample under consideration, the continuous 
change in $E_{\sigma T}$ can be approximated by a linear function. In addition, the 
abrupt change of $E_{\sigma T}$ at the phase transition must be taken into account. 
The following phenomenological description can be chosen:

\begin{equation}
\label{eq:sigmoid}
\ E_{\sigma T}= E_0+E_1 X + \frac{\Delta E}{1+e^{w(X_0-X)}}
\end{equation}

Here, $X=1/T$ and $X_0=1/T_0$ represent the reciprocal temperature in general and at 
the turning point of the sigmoid function used here to describe the width $w$ and 
magnitude $\Delta E$ of the abrupt change in $E_{\sigma T}$. Furthermore, $E_0$ and 
$E_1$ are the coefficients of the linear function to approximate the continuous change.

\begin{figure}
	\centering
	\includegraphics[width=0.45\linewidth]{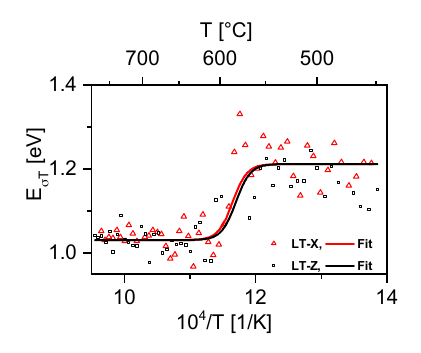}
	\includegraphics[width=0.45\linewidth]{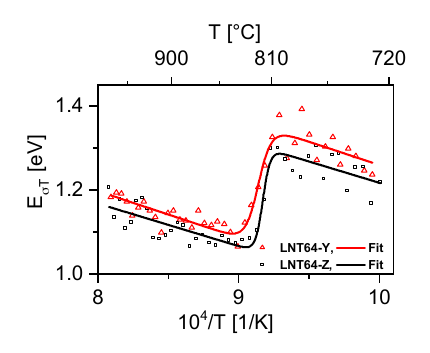}\\
	\includegraphics[width=0.45\linewidth]{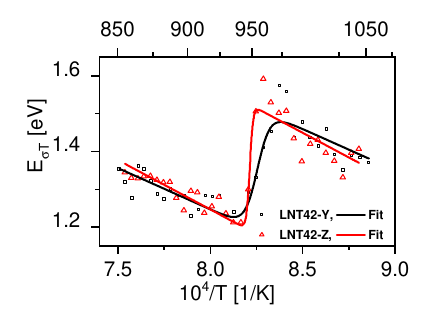}
	\includegraphics[width=0.45\linewidth]{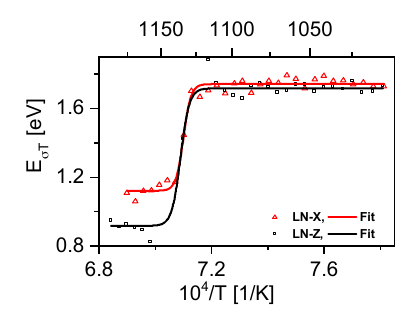}
	\caption{Representative slopes of logarithmic conductivity for samples of LNT (symbols) and fit (line) in the vicinity of the respective phase transition.}\label{fig:LNT_Fit}
\end{figure}

Since the exact shape of the abrupt change in $E_{\sigma T}$ is unknown, the sigmoid 
function chosen here is a straight forward approximation that requires the least assumptions. 
The choice of such a function is justified if a closer look is taken at $E_{\sigma T}$ in 
the vicinity of the phase transition as done in Fig.\,\ref{fig:LNT_Fit}. However, the 
relationship between reciprocal temperature and $E_{\sigma T}$ cannot be specified further 
due to scattering of the latter. Note, that the scatter results from the fact $E_{\sigma T}$ 
is a derivative.

The parameters of Eq.\,\ref{eq:sigmoid} are fitted to $E_{\sigma T}$. Beside $\Delta E$, 
$X_0$ is relevant and discussed in relation to $T_c$. Here, $w$, $E_0$ and $E_1$ enable 
fitting, but are not used further.  Figure\,\ref{fig:LNT_Fit} presents the fit of 
Eq.\,\ref{eq:sigmoid} to $E_{\sigma T}$ in the vicinity of the phase transition. The 
enlarged representation also shows no dependence on the crystal direction.

The numerical results of the fitting are summarized in Tab.\,\ref{Tab:Tc} and compared with $T_c$ from DSC and theory.

\begin{table}
\centering
\caption{The Curie temperature of LNT determined by DSC (numbers in bold), MD calculations (AIMD, SSCHA) and E$_{\sigma T}$-values (averaged numbers for both directions in bold) along with a comparison of activation energies of conductivity during phase transformation. The sample name applies for the conductivity measurements only.}	
\begin{tabular}[htbp]{@{}llllllll@{}}
\hline
\multicolumn{1}{r|}{Method} & \multicolumn{1}{l|}{DSC} & \multicolumn{1}{l|}{AIMD} & \multicolumn{1}{l|}{SSCHA} & \multicolumn{4}{c}{Conductivity} \\
\hline
\multicolumn{1}{l|}{Composition} &\multicolumn{5}{c|}
{T$_c$ [K]}&\multicolumn{1}{c|}{$\Delta$E [eV]}& Sample\\
\hline
\multirow{2}*{1.0} & \multirow{2}*{\textbf{873}} & \multirow{2}*{841} & \multirow{2}*{808} & \multirow{2}*{\textbf{870}} & 862 & 0.181 $\pm$ 0.002 & LT-X\\
&  &  & & & 877 & 0.181 $\pm$ 0.011 & LT-Z\\
\hline
\multirow{2}*{0.64}  & \multirow{2}*{\textbf{1092}} &  \multirow{2}*{-} & \multirow{2}*{-} & \multirow{2}*{\textbf{1092}} & 1094& 0.284 $\pm$ 0.025 & LNT64-Y\\
&  &  & & & 1089 & 0.252 $\pm$ 0.023 & LNT64-Z\\
\hline
\multirow{2}*{0.42}  & \multirow{2}*{\textbf{1215}} &  \multirow{2}*{-} & \multirow{2}*{-} & \multirow{2}*{\textbf{1215}} & 1218 & 0.333 $\pm$ 0.027 & LNT42-Y\\
&  &  & & & 1212  & 0.314 $\pm$ 0.028 & LNT42-Z\\
\hline
\multirow{2}*{0.0}  & \multirow{2}*{\textbf{1415}} & \multirow{2}*{1524} & \multirow{2}*{1408} & \multirow{2}*{\textbf{1410}} & 1411 & 0.621 $\pm$ 0.007 & LN-X\\
&  &  & & & 1408 & 0.800 $\pm$ 0.005 & LN-Z\\
\hline
\label{Tab:Tc}
\end{tabular}
\end{table}

\section{\label{sec:Discussion}Discussion}

The first and most remarkable result is the change in activation energy of the electrical 
conductivity upon the ferroelectric-paraelectric phase transition. The fact is discussed 
below in concert with the dependence of $\Delta E$ from composition.

\begin{figure}
	\includegraphics[width=0.6\linewidth]{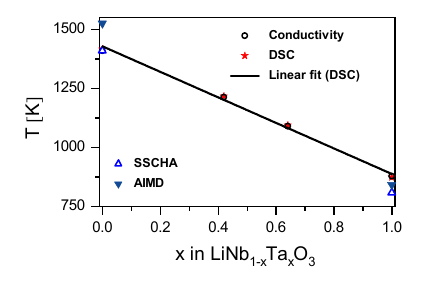}
	\centering
	\caption{Curie temperatures of LNT determined experimentally (DSC, electrical conductitity) and calculated (SSCHA, AIMD). The linear fit for DSC data corresponds to that in Fig.\,\ref{fig:Cp_Lit}.} \label{fig:Tc_DSC_Sigma}
\end{figure}

The second key result is the agreement between the (reciprocal) turning points $1/X_0$ and 
the Curie temperatures determined from $E_{\sigma T}$ and DSC data, respectively, for a given 
composition. Consequently, $1/X_0$ represents the Curie temperature, in this case, derived 
from electrical conductivity. Accordingly, $1/X_0$ is also denoted as $T_c$ in Tab.\,\ref{Tab:Tc}. 
The experimental data compared here are highlighted in bold in Tab.\,\ref{Tab:Tc}. Furthermore, 
the situation is visualized in Fig.\,\ref{fig:Tc_DSC_Sigma}. The measured data are also in reasonable agreement with the teoretical estimate of $T_c$ resulting from molecular dnymics or the SSCHA algorithm \cite{SimoMD23}.

Thirdly, the abrupt drop $\Delta E$ in $E_{\sigma T}$ at the transition from the 
ferroelectric to the paraelectric phase increases with decreasing Ta content. The 
latter goes along with the increasing temperature of the phase transition.

The discussion of these results must be done based on the mechanisms of the charge 
transport. Up to about 600\,$^\circ$C, lithium migration dominates. Above that temperature, transport via 
free polarons contributes increasingly. This contribution depends on the composition, 
and is particularly significant in Nb rich LNT \cite{Yakhnevych2023}. Consequently, the impact of the phase transition on both the ionic and electronic 
transport must be considered.

\subsection{Effect of the phase transition on the ionic transport}
The decrease of $E_{\sigma T}$ when exceeding the Curie temperature $T_c$ is the 
most evident phenomenon in case of LT. Here, in the ferroelectric and the 
paraelectric phase below and above $T_c$, respectively, ranges with constant 
$E_{\sigma T}$ exist, which can be assigned to activation energies denoted by 
$E_{AI}$ in Tab.\,\ref{Tab:Ea}. We attribute the decrease of the activation 
energy to the peculiar distribution of the Li ions on the octahedral sites, 
which are involved in the phase transition near $T_c$. The occupation of the
octahedral sites depends on the temperature, as revealed by AIMD and shown in 
Figure\,\ref{fig:Li-distr-Temp}.

\begin{figure}
	\centering
	\includegraphics[width=0.8\linewidth]{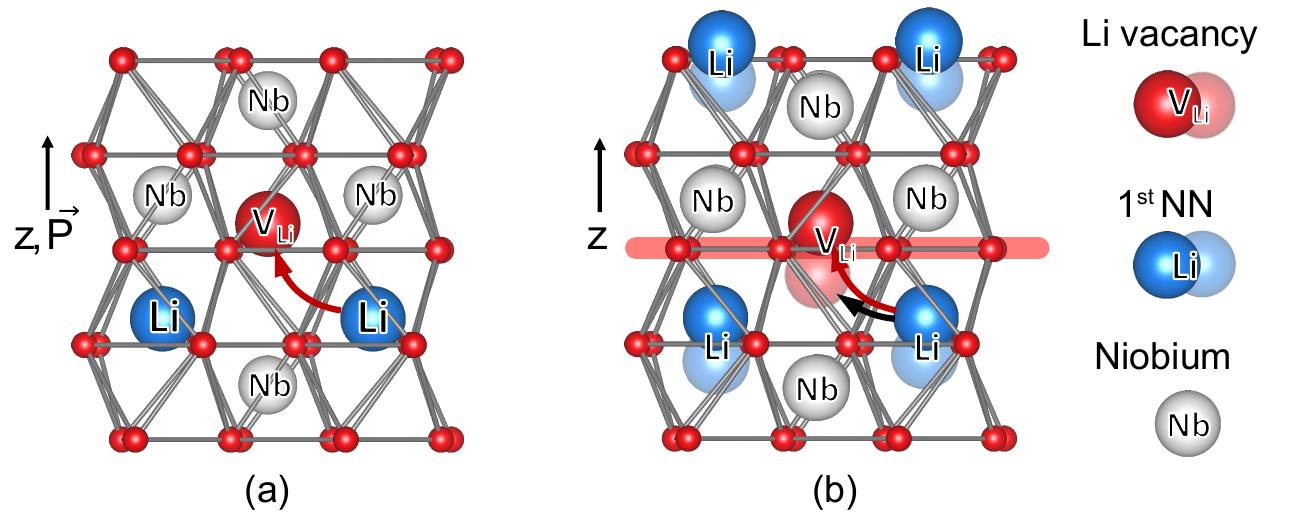}
	\caption{Position of the Li ions and vacancies in the ferroelectric phase [Ref12] (a) and
paraelectric phase with Li ions expected with a probability of 0.5 in two adjacent positions (dark and light blue) (b).}\label{fig:Li_Vacancies}
\end{figure}

Figure\,\ref{fig:Li_Vacancies}(a) shows the position of the Li ions and vacancies 
in the ferroelectric phase, as well as the preferred path (red arrow) for the Li 
migration with an activation energy of 1.29 eV \cite{Kofahl23}. In the paraelectric phase, the Li ions are expected with a probability of 0.5 in two adjacent positions, given respectively by the dark and light blue positions in Fig.\,\ref{fig:Li_Vacancies}(b) \cite{Sanna2012}. As a consequence, half of the 
moving Li vacancies no longer have to cross the repulsive oxygen plane (red line), 
resulting in a shorter transport path (black arrow). In the ferroelectric phase, 
the potential barrier has a maxium when the diffusion path leads the moving Li atoms 
between two Nb ions \cite{Kofahl23}.  We suggest that the energy barrier for the Li
diffusion in the paraelectric phase has a maximum at a similar position. However, 
the barrier height has a slightly lower average value due to the different distribution 
of the Li ions. The question of whether the specific difference of the activation energy 
of 0.181~eV below and above $T_c$ can be explained in this way has to be clarified by 
further atomistic calculations.

As long as the contribution of electrons to the electrical conductivity is low, the 
decrease in activation energy is seen as an effect of the Li transport. The condition 
is fulfilled for LT around 870\,K.  

\subsection{Effect of the phase transition on the electronic transport}
Electronic transport in LN is usually interpreted in terms of polaron hopping. A polaron is 
formed in materials with high lattice polarizability, when the lattice deformation induced by 
a charge carrier is strong enough to localize the carrier substantially at a single lattice 
site (small polaron). Charge carriers (electrons or holes, in LN and LT mostly electrons) 
localized at a regular lattice site are referred to as free small polarons, whereas they are 
referred to as bound polarons if they are localized at a lattice defect. Due to the lattice 
distortion, tunneling to equivalent lattice sites is strongly quenched, and only thermally 
activated hopping allows transport to neighboring sites.
Transport can occur in form of hopping, when an activation energy between the initial and 
final state is overcome. In the case of free polarons, thermally induced phonon excitations 
might render the electronic energies of the two sites equal, so that a transition between 
both states by tunneling is greatly facilitated. 

Differently from free polarons, the coupling to a single point defect removes the Bloch 
character from bound polarons. The binding to the defect by coulomb interaction causes a 
more pronounced distortion-related localization and stabilization. Moreover, the presence 
of the defect causes an inequivalence between the initial and the available final states of 
a hopping process. Thus, the mixture of initial and final states is asymmetric, strongly 
reducing the intensity of a charge transfer transition. 

The polaronic contribution to electronic transport is therefore expected to stem mainly 
from free polarons \cite{Schirmer09}, which might become rather mobile at higher 
temperatures, at which thermal fluctuations of the lattice generate configurations in 
which the electron energies at the initial and the final sites are equal.

Another aspect that could contribute to the decrease of the activation energy during 
the transition from the para- to the ferroelectric phase is the narrowing of the band 
gap accompanying the transition. Many-body perturbation theory indicates a decrease of 
the band gap energy by 0.5\,eV for LN, for example \cite{Thierfelder10}. This decrease 
supports the excitation of electrons, which in turn contribute to polaron transport.

\subsection{Comparison of experiments and atomistic calculations}

\begin{figure}
\Centering
\includegraphics[width=0.6\linewidth]{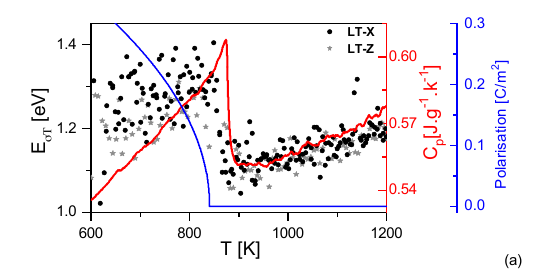}
\includegraphics[width=0.6\linewidth]{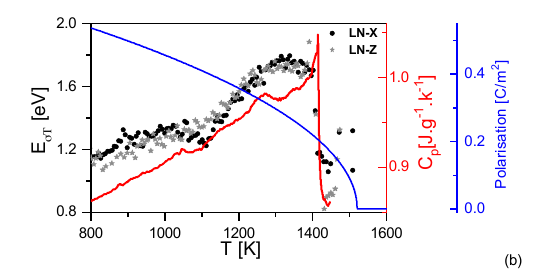}
\caption{Calculated spontaneous polarisation as a function of temperature (blue line), measured electrical conductivity (symbol) and measured thermal capacity at constant pressure $C_p(T)$ (red line) of LiTaO$_3$ (a) and LiNbO$_3$ (b). }\label{fig:Sigma_MD_DSC}
\end{figure}

Temperature dependent DSC data and the slope of the electrical conductivity $E_{\sigma T}$ of LN and LT are compared in Fig.\,\ref{fig:Sigma_MD_DSC}. The Curie temperatures determined experimentally by both methods agree very well, see also Tab.\,\ref{Tab:Tc}. This statements holds true for LNT crystals, too. Consequently, the evaluation of the electrical conductivity is an elegant approach to determine the Curie temperature as the turning point of the fitted sigmoid function, respective $T_c$, is clearly visible in the measurements carried out here.

Furthermore, Fig.\,\ref{fig:Sigma_MD_DSC} encloses the temperature dependent polarization of LN and LT as calculated from the AIMD trajectories \cite{SimoMD23}. The AIMD determined onset of the 
spontaneous polarization at 841\,K and 1524\,K for LT and LN, respectively, corresponds roughly to the range at which the activation energy of the electrical conductivity changes  from the value of the paraelectric to the value of the ferroelectric phase. In the case of LT, the AIMD calculations result in $T_c$ that matches the experimental values reasonably well. Here, the activation energy of the electrical conductivity decreases between 830\,K and 890\,K. The related change of $C_p$ is observed between 873\,K and 890\,K. In the case of LN, however, the AIMD calculations overestimate $T_c$, resulting in a small deviation from the experimental values. Here, the changes of $E_{\sigma T}$ and $C_p$ are observed between 1390\,K and 1443\,K and between 1415\,K and 1443\,K, respectively. Based on the calculated and measured data, it can be concluded that the phase transition is a continuous process that affects the properties of the materials below the critical temperature $T_c$.

\section{\label{sec:5. Conclusion }Conclusions}

An investigation of the ferroelectric to paraelectric phase transition of LiNb$_{1-x}$Ta$_x$O$_3$ is presented, which revealed a linear dependence of the phase transition temperature near $T_c$ on the composition $x$. Beside differential scanning calorimetry data, the slope of the electrical conductivity is evaluated. The latter enables to extract $T_c$. Consequently, the measurement of the electrical conductivity is an elegant approach to determine $T_c$. Further, the conductivity slope reveals a significant change in activation energy during the transition. The observation is, on the one hand, attributed to the peculiar distribution of the Li ions on the octahedral sites which affects the Li ion transport. On the other hand, with rising temperatures and, in particular, for low Ta content $x$, the transport of polarons contributes increasingly to the total conductivity. Here, lattice distortion is expected to affect the hopping of polarons. The latter might become rather mobile at higher temperatures and, consequently, thermal fluctuations of the lattice generate configurations in which the electron energies at the initial and the final sites are equal, thus facilitating polaron hopping. Furthermore, molecular dynamics calculations within the framework of the density functional theory show that substantial cation displacements occur below the transition temperature for both LiNbO$_3$ and LiTaO$_3$, and presumably LiNb$_{1-x}$Ta$_x$O$_3$. These displacements might, therefore, affect the properties of the materials already below $T_c$.

\medskip
\textbf{Acknowledgements} \par 
The Deutsche Forschungsgemeinschaft (DFG, German Research Foundation) is 
gratefully acknowledged for financial support within the research unit FOR5044 
(GA 2403/7-1, FR 1301/40-1, FR 1301/42-1, SA 1948/2-1). In addition, the authors thank the Research Center Energy Storage Technologies (Forschungszentrum Energiespeichertechnologien) for support. Calculations for this research were conducted on the Lichtenberg high-performance computer of the TU Darmstadt and at the 
H\"ochstleistungsrechenzentrum Stuttgart (HLRS). The authors furthermore acknowledge 
the computational resources provided by the HPC Core Facility and the HRZ of the 
Justus-Liebig-Universit\"at Gie{\ss}en.

\medskip

%

\bibliography{literature_LNT}
\bibliographystyle{abbrv}

\end{document}